\documentclass[%
 reprint,
superscriptaddress,
 amsmath,amssymb,
 aps,
]{revtex4-2}

\usepackage{graphicx}% Include figure files
\usepackage{dcolumn}% Align table columns on decimal point
\usepackage{bm}% bold math
\usepackage{mathtools}
\usepackage{gensymb}
\usepackage{amsmath}
\usepackage{qcircuit}
\usepackage[utf8]{inputenc}
\usepackage{color,soul}
\usepackage{float}

\DeclarePairedDelimiter\bra{\langle}{\rvert}
\DeclarePairedDelimiter\ket{\lvert}{\rangle}

\begin{document}

\title{Refrigeration with Indefinite Causal Orders \\ on a Cloud Quantum Computer}% Force line breaks with \\

\author{David Felce}
% \altaffiliation[Also at ]{Physics Department, XYZ University.}%Lines break automatically or can be forced with \\
\email{david.felce@physics.ox.ac.uk}
\affiliation{Clarendon Laboratory, Department of Physics, University of Oxford, England}

\author{Vlatko Vedral}
\affiliation{Clarendon Laboratory, Department of Physics, University of Oxford, England}
\affiliation{Centre for Quantum Technologies, National University of Singapore, Block S15, 3 Science Drive 2, Singapore}
\affiliation{Department of Physics, National University of Singapore, Science Drive 3, Blk S12, Level 2, Singapore 1175512}

\author{Felix Tennie}
\affiliation{Clarendon Laboratory, Department of Physics, University of Oxford, England}

\begin{abstract}
We demonstrate non-classical cooling on the IBMq cloud quantum computer. We implement a recently proposed refrigeration protocol which relies upon indefinite causal order for its quantum advantage. We use quantum channels which, when used in a well-defined order, are useless for refrigeration. We are able to use them for refrigeration, however, by applying them in a superposition of different orders. Our protocol is by nature relatively robust to noise, and so can be implemented on this noisy platform. As far as the authors are aware, this is the first example of cloud quantum refrigeration.
\end{abstract}

%\keywords{Suggested keywords}%Use showkeys class option if keyword
                              %display desired
\maketitle

%\tableofcontents

\begin{figure}[h]
 \includegraphics[width=\linewidth]{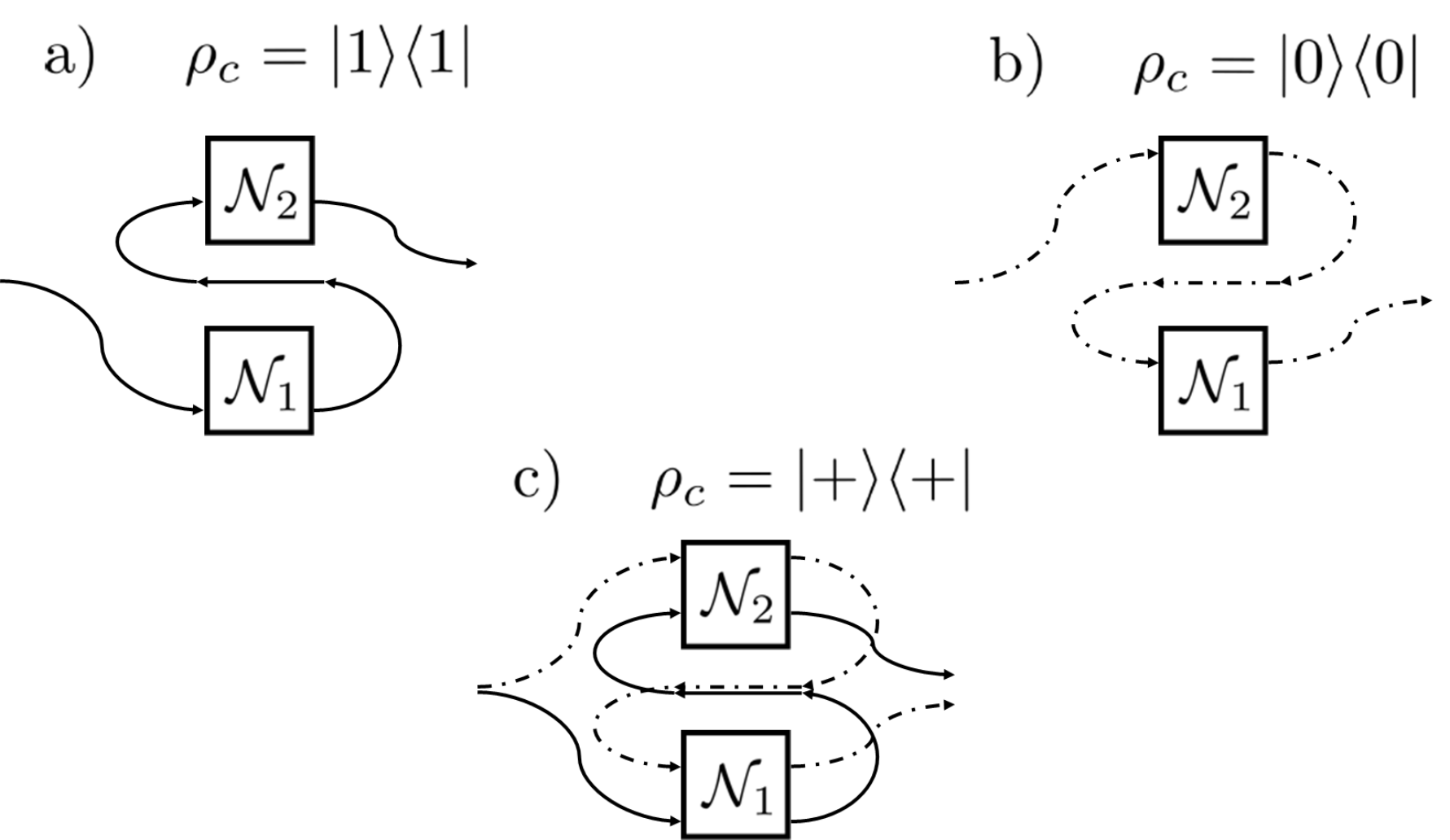}
 \caption{(a) and (b) illustrate channels $\mathcal{N}_1$ and $\mathcal{N}_2$ placed in a definite order, corresponding to the control qubit being in state $\ket{1}\bra{1}$ and $\ket{0}\bra{0}$ respectively. In (c) the quantum SWITCH places the channels in a superposition of causal orders. It entangles the order of the two channels with the state of the control qubit, in this case  $\ket{+}\bra{+}$. (c.f. equations \ref{eq:SN1N2},\ref{eq:Wij})} 
 \label{fig:QSWITCH}
\end{figure}

\paragraph*{Introduction---}
Events which take place in a quantum superposition of different orders are said to exhibit an indefinite causal order (ICO). 
The quantum SWITCH \citep{PhysRevA.88.022318} is one particular form of indefinite causal structure, in which the order of operation of two or more quantum channels is determined by the state a control qubit  (Fig.~\ref{fig:QSWITCH}). There is a growing literature discussing the advantage that the quantum SWITCH can provide in quantum computation \citep{PhysRevA.86.040301,Colnaghi2012,PhysRevLett.113.250402},  communication \citep{PhysRevLett.120.120502,2018arXiv180906655S,chiribella2018indefinite,2019arXiv190201807P}, and metrology \citep{2018arXiv181207508M,zhao2019quantum,DBLP:journals/qip/Frey19,PhysRevA.103.032615}.
 There have even been recent experiments \citep{Procopio2015,rubino2017,PhysRevLett.121.090503,goswami2018communicating,PhysRevLett.122.120504,PhysRevLett.124.030502} realizing indefinite causal orders in the laboratory setting and using them to perform ICO communication protocols.  \par
A recent work \citep{ICOFridgeTheory} demonstrated that the quantum SWITCH could be used to allow quantum refrigeration using thermalizing channels that would ordinarily be useless for this task. There have been experiments aiming to implement this protocol in nuclear magnetic resonance \citep{nie2020experimental}, and photonic \citep{cao2021experimental} systems. In this work, we implement the quantum refrigeration protocol using superconducting qubits, hosted on a cloud based server which is open to all users of the internet. \par
Quantum refrigeration via the quantum SWITCH can be described within the framework of quantum channels and their Kraus operators \cite{NielsenChuang}. Mathematically, the quantum SWITCH of two quantum channels is itself a quantum channel. The output of the quantum SWITCH of two channels $\mathcal{N}_1$ and $\mathcal{N}_2$ takes the form

\begin{equation} 
S(\mathcal{N}_1,\mathcal{N}_2)(\rho_c\otimes\rho) = \sum_{i,j}W_{ij}(\rho_c\otimes\rho)W^\dagger_{ij},
\label{eq:SN1N2}
\end{equation}

where

\begin{equation} 
W_{ij} = \ket{0}\bra{0}_c \otimes K^{(1)}_iK^{(2)}_j + \ket{1}\bra{1}_c \otimes K^{(2)}_jK^{(1)}_i,
\label{eq:Wij}
\end{equation}
are the Kraus operators of the composite channel generated by the quantum SWITCH. These operators are comprised of the Kraus operators $K$ of the original channel, applied in an order which depends on the state of the control qubit. $\rho_c$ and $\rho$ denote the density operators of the control and target systems respectively.

It has been shown \citep{ICOFridgeTheory} that the quantum SWITCH of two thermalizing channels displays some remarkable properties. For initial $\rho_c = \ket{+}\bra{+}_c$, the output of the quantum SWITCH of these channels, after measurement of the control qubit in the $\ket{\pm}_c$ basis, reads

\begin{equation} 
_c\bra{\pm}S(\mathcal{N}^T,\mathcal{N}^T)(\rho_c\otimes\rho)\ket{\pm}_c = \frac{T}{2} \pm \frac{1}{2} T\rho T,
\label{eq:CondICO}
\end{equation}
where $T$ is the density operator representing the thermal state which is the output of the thermalizing channel $\mathcal{N}^T$. Note that if the channels are performed in any classically determined or random order, then the output must be $T$. We see however, that this expected outcome is not obtained when the channels are applied in an indefinite causal order. Instead, the output state retains some dependence on the input, and in general, the output state will not have the same effective temperature as that of the state $T$. \par
These features can be used to create a thermodynamic refrigeration cycle which transfers heat from cold to hot reservoirs, by thermalizing a working qubit with those reservoirs in an indefinite causal order. The target system, henceforth called the working system, starts at the temperature of the two cold reservoirs. It is acted on by the quantum SWITCH of two thermalizing channels, one for each reservoir. The control qubit is then measured in the $\ket{\pm}_c$ basis. If the result of the measurement yields $\ket{+}\bra{+}_c$, then the working system has been cooled down. It is then thermalized classically with the cold reservoirs. If the result yields $\ket{-}\bra{-}_c$ then the working system is first thermalized with the hot reservoir, and then finally with the cold reservoir. The net result of this cycle is that heat is transferred from the cold to the hot reservoir.

\paragraph*{Methods---}
Our aim was to implement this refrigeration protocol in a physical system. We run the experiment on the cloud quantum computing platform provided by IBM Quantum. Specifically, we used the \textit{ibmq}\_\textit{5}\_\textit{yorktown} backend, which is one of the IBM Quantum Canary Processors. The processor consists of 5 superconducting qubits with high connectivity and relatively high fidelity, and so is ideally suited for our purpose.
It is possible to construct a Quantum Switch of thermalizing channels using a unitary quantum circuit which acts on the control and target systems, plus environments \citep{ICOFridgeTheory}:

\begin{align} 
\centering
\Qcircuit @C=1.5em @R=2em {
\lstick{\rho_c} & \ctrl{2} & \ctrl{3} & \ctrlo{3} & \ctrlo{2} &  \qw & & \dstick{\big\}S(\rho_{in} \otimes\rho_c)} \\
\lstick{\rho_{in}} & \qswap & \qswap & \qswap & \qswap &  \qw \\
\lstick{T} & \qswap & \qw & \qw & \qswap &  \qw \\
\lstick{T} & \qw & \qswap & \qswap & \qw &  \qw 
}
\nonumber
\end{align}
\medskip

\noindent Here the order in which the swaps occur is determined by the state of the control qubit. After this circuit the marginal state of the upper two qubits is identical to $S(\mathcal{N}^T,\mathcal{N}^T)(\rho\otimes\rho_c)$ in equation (\ref{eq:CondICO}). Although in principle there are an infinite number of circuits, with different environments, which yield this output, this particular representation is helpful, because the environments (two bottom inputs $T$) can be thought of as qubits randomly drawn from two reservoirs which are  thermal baths of qubits each with thermal state $T$. Any heat flow to or from these qubits thus implies a heat flow to or from a reservoir. It is then clear that the individual swaps performed are unitary extensions of the individual quantum thermalizing channels we wish to implement in the quantum SWITCH, so we implement the correct channels in an order determined by the control qubit. \par
We first notice that the circuit above can be simplified using circuit identities to become:

\begin{align} 
\centering
\Qcircuit @C=1.5em @R=2em {
\lstick{\rho_c} & \ctrl{2} &\qw & \ctrlo{2} &  \qw & & \dstick{\big\}S(\rho_{in} \otimes\rho_c)} \\
\lstick{\rho_{in}} & \qswap & \qswap \qwx[2] & \qswap &  \qw \\
\lstick{T} & \qswap & \qw &  \qswap &  \qw \\
\lstick{T} & \qw & \qswap & \qw & \qw 
}
\nonumber
\end{align}
\medskip

The (controlled) swap gates can be decomposed into toffoli and C-NOT gates. The circuit is then transpiled using the native IBMq transpiler for the allowed operations and connectivity of the particular backend which has been selected for the experiment. The depth of the final transpiled circuit is 31. The circuit consists of 51 gates, of which 19 are two qubit gates. These two qubit gates have a lower fidelity, so care is taken to minimize the number of such gates in the circuit. \par
\begin{figure*}[htbp]
    \includegraphics[width=\textwidth]{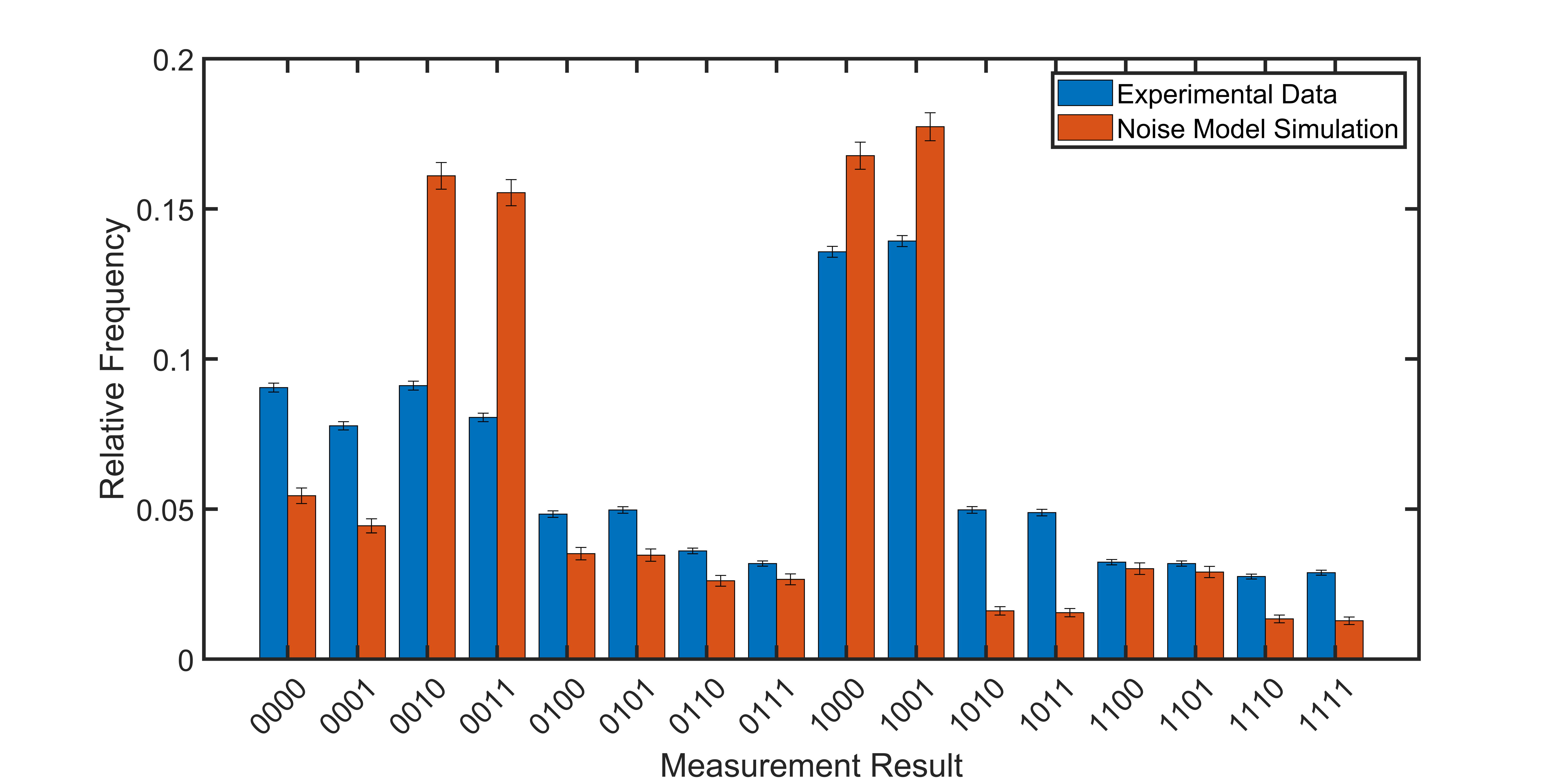}
    \caption{Histogram of the output measurement results for a particular input computational basis state (in this case work qubit `1', reservoir $A$ `0' and reservoir $B$ `0'). The measurement outcome labels on the horizontal axis have bits which correspond to, from least to most significant bit: control qubit, reservoir $A$, work qubit, reservoir $B$. \textit{Left, Blue} -- Experimental Result. \textit{Right, Red} -- QASM Noise Model Prediction. It is easily observed that the result is not compatible with the noise model within statistical error. This indicates the presence of correlated noise which is not accounted for in the model.}
    \label{fig:histogram}
\end{figure*}

The control qubit is initialised in the state $\ket{+}_c$. The refrigeration protocol requires that the work and reservoir qubits are initialized in thermal mixed states, all at the same inverse temperature, $\beta_\textrm{in}$. In order to achieve this, we run the circuit many times with each possible computational basis state for the thermal qubits. One example outcome for a particular input basis state is shown in Fig.~\ref{fig:histogram} (Left). The desired input mixed state is created by randomly selecting runs with a weighting for each input state that depends on the desired input state. Simply by including a fifth qubit representing a hot reservoir, and relabelling the qubits depending on the measurement result of the control qubit, we can then reproduce the refrigeration cycle presented in \citep{ICOFridgeTheory}.

\begin{figure*}[t]
    \includegraphics[width=.49\textwidth]{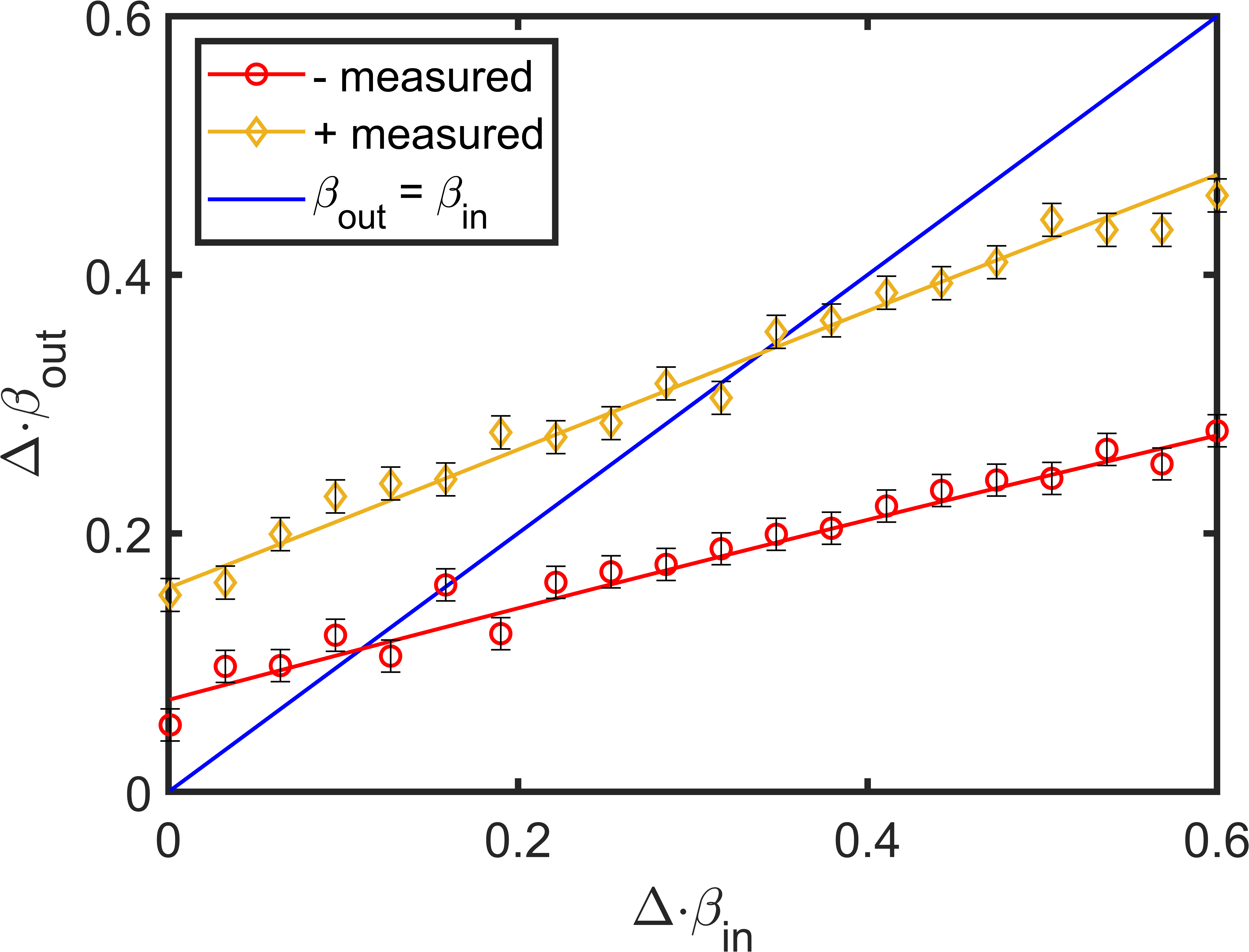}\hfill
    \includegraphics[width=.49\textwidth]{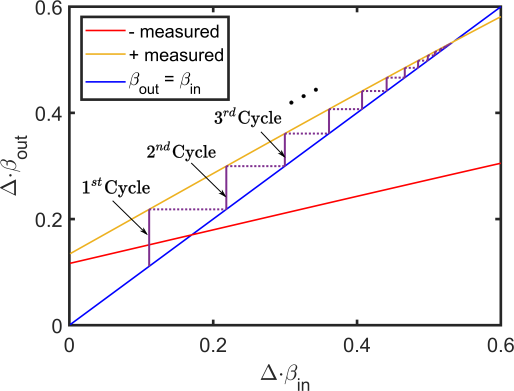}
    \\[\smallskipamount]
    \includegraphics[width=.49\textwidth]{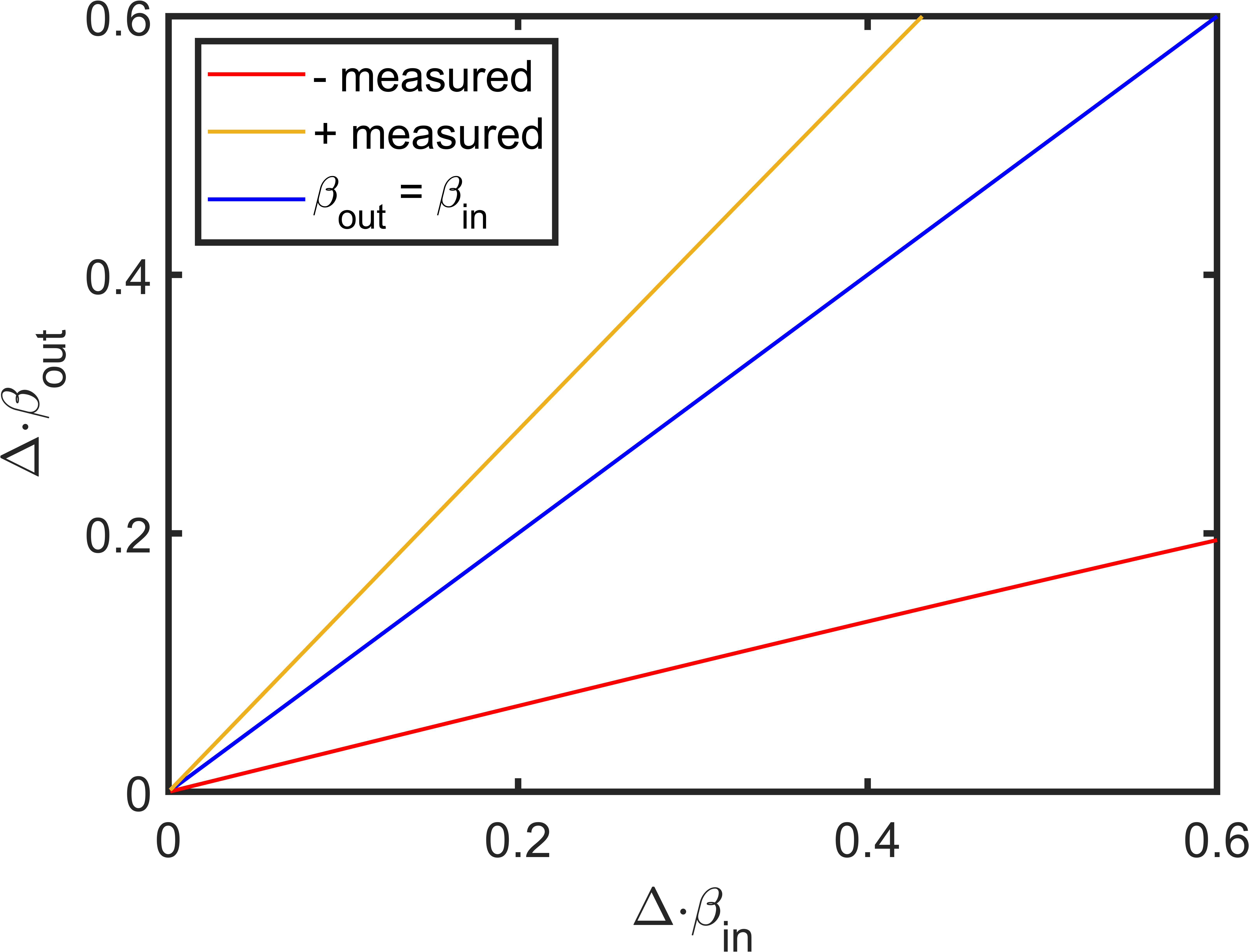}\hfill
    \includegraphics[width=.49\textwidth]{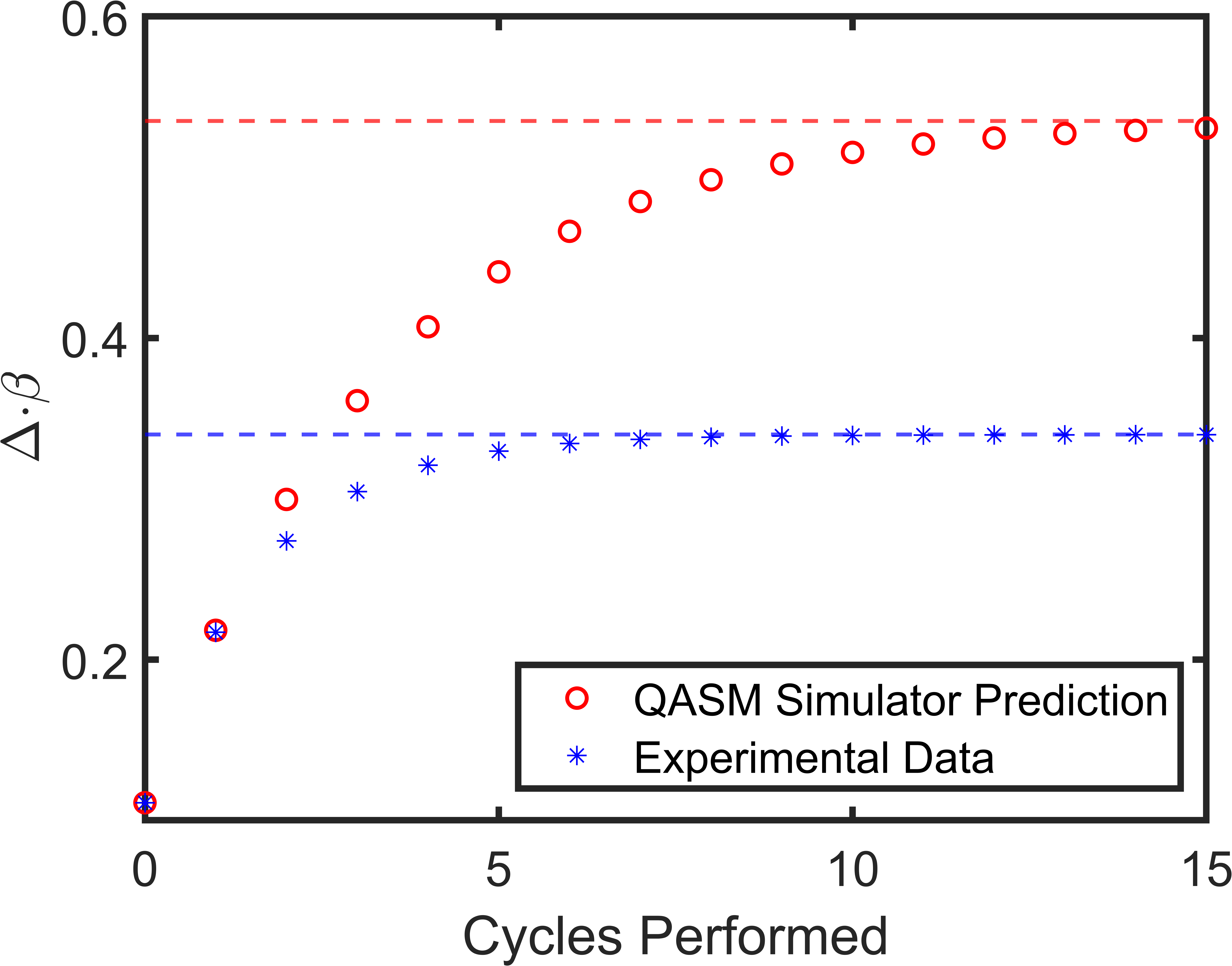}
    \caption{Dimensionless inverse temperature of the working qubit after thermalization in an indefinite causal order, plotted as a function of input temperature. The blue line represents the outcome in the classical case - the channels are performed in a definite causal order and so there is no change in temperature. The yellow line is the output temperature when the control qubit is measured to be $|+\rangle$. In this case, for all input temperatures, the output is cooler than the case when the control is found in the state $|-\rangle$ (red line). \textit{Top Left} -- Experimental Result. \textit{Top Right} -- QASM noise model simulation with IBMq backend calibration data. Solid and dashed line represents ideal temperature trajectory over multiple cycles with measurement result $|+\rangle$. \textit{Bottom Left} -- Noiseless theoretical prediction. \textit{Bottom Right} -- Dimensionless inverse temperature as a function of the number of consecutive cycles with measurement result $|+\rangle$. The temperature converges to an asymptote (blue dashed line for experiment, red dashed line for noise model simulation).}
    \label{fig:+-}
\end{figure*}

\paragraph*{Results---}
Our results demonstrate that non-classical cooling has been achieved. For a range of input temperatures, we find that the reservoir and working qubits have been cooled below their common input temperature. As the theory predicts, we find that the temperature of the thermal qubits is correlated with the state of the control qubit - something that is not consistent with the channels being applied in any classical order, thus demonstrating the crucial role of indefinite causal order in the cooling process. For a given range of input temperatures, we observe that the thermal qubits are cooled below the input temperature when the control qubit is found in the state $|+\rangle_c$, and are heated above the input temperature when it is found in state $|-\rangle_c$, as shown in Fig.~\ref{fig:+-}. This is the prediction of the theory and the proof that quantum refrigeration has been performed. \par
In the noiseless case, the lower bound on the temperature to which the qubits can be cooled by the protocol is absolute zero. Our results, however, indicate that there is a temperature, below which the qubits cannot be cooled - this is a result of the inherent temperature of the noise associated with the operations performed. This temperature bound can be read off the graph in Fig.~\ref{fig:+-} (Top Left), from the point where the line representing the output temperature when $\ket{+}_c$ is measured crosses the line $\beta_\textrm{out} = \beta_\textrm{in}$. For the IBM Yorktown backend that was used, the inverse temperature bound was given by $\Delta\cdot\beta \approx 0.3$, where $\Delta$ is the energy gap of the qubits. It is also possible to infer from our results the minimum number of cycles necessary to cool all the thermal qubits by a certain temperature (Fig. \ref{fig:+-}, Top Left), in the ideal case where $|+\rangle_c$ is measured each time (and thus no thermalization with a hot reservoir is required). This is a lower bound on the number of cycles which must be performed. More generally, the expected number of cycles depends on $p_+$ and $p_-$ (the probability of measuring $|+\rangle_c$ and $|-\rangle_c$ respectively). For the relevant temperature ranges, it is reasonable to take $p_+ \approx p_- \approx 1/2$. Then, the expected number of cycles required will be approximately $2^n$, where n is the number of steps required in the ideal case.

%perhaps, one could dash the horizontal lines to make sure the reader does not confuse them with actual state evolution?

\paragraph*{Noise Considerations---}
There are two unexpected features of the data presented in Fig.~\ref{fig:+-} that warrant further discussion. Both of these features are manifest when the input inverse temperature is very low ($\beta_\textrm{in} \rightarrow 0$). First, we observe that the experimental data (Top Left) shows that for low $\beta_\textrm{in}$ the temperature of the working system is decreased for both possible measurements of the control qubit. Also, the temperature depends on the result of this measurement, whereas in the theoretical model (Top Right), for $\beta_\textrm{in} = 0$, we have $\beta_\textrm{out} = 0$, for both $|+\rangle_c$ and $|-\rangle_c$ measurements. The first feature of the data can be understood with reference to a very simple model of the noise associated with the operations of the computer. In this model, we assume that there are two possible outcomes of the process. With some probability the computer operates perfectly and gives the theoretically expected output for a given input temperature, $\rho_{th}(\beta_\textrm{in})$, otherwise, the output is a density matrix $\rho_{fail}$. We calculate the `success probability' ($p_{suc}$) by multiplying together the fidelities (from the backend calibration data) of each of the quantum gates and measurements performed. The output then takes the form

\begin{equation} 
\rho_{out } = p_{suc}\rho_{th}(\beta_\textrm{in}) + (1-p_{suc})\rho_{fail},
\label{eq:noisemodel}
\end{equation}

The form of $\rho_{fail}$ is unknown (the error processes can be represented by some combination of depolarizing and dephasing channels), but since the off-diagonal elements are unimportant, there is only one parameter relevant to the model: the effective temperature of the state. Since this effective temperature will be finite, and $p_{suc}$ is less than unity, $\beta_\textrm{out} = 0$ can never be achieved, even when $\beta_\textrm{in} = 0$. However, this  model does not explain why, when $\beta_\textrm{in}$ is very low, $\beta_\textrm{out}$ depends on the measurement results ($|+\rangle_c$ or $|-\rangle_c$) of the control qubit. In fact, the model above (Eq.~\ref{eq:noisemodel}) predicts that there will be no splitting in output temperature when $\beta_\textrm{in} = 0$. \par
A simple modification to the model can explain this second feature of the data. Let us assume that the input states of the process are not perfectly prepared, but rather are prepared correctly with a certain probability, and otherwise are initialized in some thermal state of unknown temperature. Thus, the effective temperature of the input states cannot be controlled precisely. It is therefore clear that the input states will never have a temperature of exactly $\beta_\textrm{in} = 0$. Therefore $\rho_{th}$ in Eq.~(\ref{eq:noisemodel}) must be replaced with 
\begin{equation}
\rho_{th}'(\beta_\textrm{in}) = p'_{suc}\rho_{th}(\beta_\textrm{in}) + (1-p'_{suc})\rho_{th}(\beta_\textrm{fail}), 
\nonumber
\end{equation}
where $p'_{suc}$ is the probability of a faithful initialization and $\beta_\textrm{fail}$ is the temperature of the input when the initialization fails. Since $\beta_\textrm{fail}$ is a finite temperature, with $\rho_{th}'$ we still expect a splitting of $\beta_\textrm{out}$ for the $|+\rangle_c$ and $|-\rangle_c$ measurement results. \par
A more sophisticated noise model, provided by IBM Quantum, with parameters specific to the quantum backend used, gives a more accurate picture of the noise processes which occur. This model includes sources of error such as single qubit pauli errors, readout errors and errors associated with C-NOT operations. We provide a sample histogram (Fig.~\ref{fig:histogram}) for comparison between the measured data (blue) and this model's prediction (red). These data arise from a particular input computational basis state of the thermal qubits. Still, there is a deviation of the measured data, even from this comprehensive noise model, which cannot be resolved within the bounds of sampling errors. This inaccuracy may arise because, either the calibration data were not accurate for the time that the experiment took place, or because of unknown and unaccounted sources of correlated noise.

\paragraph*{Discussion---}
As far as the authors are aware, this work constitutes the first thermodynamic protocol run on a programmable quantum computer. We have shown that current quantum processors can be used to successfully implement such protocols, and provide a new test bed for quantum thermodynamics. \par
The implementation of the quantum SWITCH on a quantum computer allows for easy generalization and extension of any ICO protocol. For example, it is possible to choose any desired input state of the control qubit, and any channels to place in a superpostion of orders. It is also possible to straightforwardly extend the SWITCH to $N>2$ channels and orders. This permits a full investigation of the advantage provided by the quantum SWITCH. \par
Our experiment is a proof of principle of the indefinite causal order refrigeration cycle. We have provided a very simple model of the noise which explains the major features of the observed data, as well as a more sophisticated one, capable of reproducing the key features of the data. The successful implementation of the cycle in a noisy system proves the robustness of the protocol, and indicates that other ICO protocols need not necessarily be rendered impotent on account of noise. Further understanding the impact of noise on ICO-enhanced tasks is essential to evaluating the full potential of the quantum SWITCH for applications not limited to thermodynamics, but also including communication and metrology. Our results are a promising indicator for the robustness of ICO protocols to noise. \par
Our work motivates further research on the limitations of the quantum SWITCH in assisting thermodynamic tasks, and the search for other, perhaps more efficient, (refrigeration) protocols. One open question is whether it is possible to realise an ICO thermodynamic cycle whose efficiency can in principle approach Carnot efficiency. The authors suspect this might be possible using a modified version of the cycle implemented here. \par
We have shown that a thermodynamic ICO protocol can be implemented using current technology. The search is on for more such protocols, and the prospect for their physical realization is very encouraging.

\bigskip

%\begin{list}
% x   \item Implementation in a noisy system proves robustness of the refrigeration protocol.
% x   \item Effective noise model explains major features of the observed data.
% z   \item first cloud quantum computing based thermo protocol?
%   \item Quantum signature of refrigeration process
% x \item Understanding the impact of noise on ICO-enhanced tasks essential to evaluate full potential of the quantum switch, i.e. might also be relevant to other applications in communication, etc.
% x \item next steps: our paper motivates further research on the limitations of quantum swtich in assisting thermo tasks;
%\end{list}

\paragraph*{Acknowledgements---}
The authors would like to thank Benjamin Yadin and Niel de Beaudrap for helpful discussions and comments. We acknowledge the use of IBM Quantum services for this work. The views expressed are those of the authors, and do not reflect the official policy or position of IBM or the IBM Quantum team. DF is supported by the EPSRC (UK) and by M squared. VV thanks the National Research Foundation, Prime Minister’s Office, Singapore, under its Competitive Research Programme (CRP Award No. NRF- CRP14-2014-02) and administered by Centre for Quantum Technologies, National University of Singapore. FT is supported by the Royal Society.

\bibliography{references}% Produces the bibliography via BibTeX.

%apsrev4-2.bst 2019-01-14 (MD) hand-edited version of apsrev4-1.bst
%Control: key (0)
%Control: author (8) initials jnrlst
%Control: editor formatted (1) identically to author
%Control: production of article title (0) allowed
%Control: page (0) single
%Control: year (1) truncated
%Control: production of eprint (0) enabled
\begin{thebibliography}{22}%
\makeatletter
\providecommand \@ifxundefined [1]{%
 \@ifx{#1\undefined}
}%
\providecommand \@ifnum [1]{%
 \ifnum #1\expandafter \@firstoftwo
 \else \expandafter \@secondoftwo
 \fi
}%
\providecommand \@ifx [1]{%
 \ifx #1\expandafter \@firstoftwo
 \else \expandafter \@secondoftwo
 \fi
}%
\providecommand \natexlab [1]{#1}%
\providecommand \enquote  [1]{``#1''}%
\providecommand \bibnamefont  [1]{#1}%
\providecommand \bibfnamefont [1]{#1}%
\providecommand \citenamefont [1]{#1}%
\providecommand \href@noop [0]{\@secondoftwo}%
\providecommand \href [0]{\begingroup \@sanitize@url \@href}%
\providecommand \@href[1]{\@@startlink{#1}\@@href}%
\providecommand \@@href[1]{\endgroup#1\@@endlink}%
\providecommand \@sanitize@url [0]{\catcode `\\12\catcode `\$12\catcode
  `\&12\catcode `\#12\catcode `\^12\catcode `\_12\catcode `\%12\relax}%
\providecommand \@@startlink[1]{}%
\providecommand \@@endlink[0]{}%
\providecommand \url  [0]{\begingroup\@sanitize@url \@url }%
\providecommand \@url [1]{\endgroup\@href {#1}{\urlprefix }}%
\providecommand \urlprefix  [0]{URL }%
\providecommand \Eprint [0]{\href }%
\providecommand \doibase [0]{https://doi.org/}%
\providecommand \selectlanguage [0]{\@gobble}%
\providecommand \bibinfo  [0]{\@secondoftwo}%
\providecommand \bibfield  [0]{\@secondoftwo}%
\providecommand \translation [1]{[#1]}%
\providecommand \BibitemOpen [0]{}%
\providecommand \bibitemStop [0]{}%
\providecommand \bibitemNoStop [0]{.\EOS\space}%
\providecommand \EOS [0]{\spacefactor3000\relax}%
\providecommand \BibitemShut  [1]{\csname bibitem#1\endcsname}%
\let\auto@bib@innerbib\@empty
%</preamble>
\bibitem [{\citenamefont {Chiribella}\ \emph {et~al.}(2013)\citenamefont
  {Chiribella}, \citenamefont {D'Ariano}, \citenamefont {Perinotti},\ and\
  \citenamefont {Valiron}}]{PhysRevA.88.022318}%
  \BibitemOpen
  \bibfield  {author} {\bibinfo {author} {\bibfnamefont {G.}~\bibnamefont
  {Chiribella}}, \bibinfo {author} {\bibfnamefont {G.~M.}\ \bibnamefont
  {D'Ariano}}, \bibinfo {author} {\bibfnamefont {P.}~\bibnamefont
  {Perinotti}},\ and\ \bibinfo {author} {\bibfnamefont {B.}~\bibnamefont
  {Valiron}},\ }\bibfield  {title} {\bibinfo {title} {Quantum computations
  without definite causal structure},\ }\href
  {https://doi.org/10.1103/PhysRevA.88.022318} {\bibfield  {journal} {\bibinfo
  {journal} {Phys. Rev. A}\ }\textbf {\bibinfo {volume} {88}},\ \bibinfo
  {pages} {022318} (\bibinfo {year} {2013})}\BibitemShut {NoStop}%
\bibitem [{\citenamefont {Chiribella}(2012)}]{PhysRevA.86.040301}%
  \BibitemOpen
  \bibfield  {author} {\bibinfo {author} {\bibfnamefont {G.}~\bibnamefont
  {Chiribella}},\ }\bibfield  {title} {\bibinfo {title} {Perfect discrimination
  of no-signalling channels via quantum superposition of causal structures},\
  }\href {https://doi.org/10.1103/PhysRevA.86.040301} {\bibfield  {journal}
  {\bibinfo  {journal} {Phys. Rev. A}\ }\textbf {\bibinfo {volume} {86}},\
  \bibinfo {pages} {040301(R)} (\bibinfo {year} {2012})}\BibitemShut {NoStop}%
\bibitem [{\citenamefont {Colnaghi}\ \emph {et~al.}(2012)\citenamefont
  {Colnaghi}, \citenamefont {D'Ariano}, \citenamefont {Facchini},\ and\
  \citenamefont {Perinotti}}]{Colnaghi2012}%
  \BibitemOpen
  \bibfield  {author} {\bibinfo {author} {\bibfnamefont {T.}~\bibnamefont
  {Colnaghi}}, \bibinfo {author} {\bibfnamefont {G.~M.}\ \bibnamefont
  {D'Ariano}}, \bibinfo {author} {\bibfnamefont {S.}~\bibnamefont {Facchini}},\
  and\ \bibinfo {author} {\bibfnamefont {P.}~\bibnamefont {Perinotti}},\
  }\bibfield  {title} {\bibinfo {title} {Quantum computation with programmable
  connections between gates},\ }\href
  {https://doi.org/10.1016/j.physleta.2012.08.028} {\bibfield  {journal}
  {\bibinfo  {journal} {Physics Letters A}\ }\textbf {\bibinfo {volume}
  {376}},\ \bibinfo {pages} {2940} (\bibinfo {year} {2012})}\BibitemShut
  {NoStop}%
\bibitem [{\citenamefont {Ara\'ujo}\ \emph {et~al.}(2014)\citenamefont
  {Ara\'ujo}, \citenamefont {Costa},\ and\ \citenamefont
  {Brukner}}]{PhysRevLett.113.250402}%
  \BibitemOpen
  \bibfield  {author} {\bibinfo {author} {\bibfnamefont {M.}~\bibnamefont
  {Ara\'ujo}}, \bibinfo {author} {\bibfnamefont {F.}~\bibnamefont {Costa}},\
  and\ \bibinfo {author} {\bibfnamefont {{\v C}.}~\bibnamefont {Brukner}},\
  }\bibfield  {title} {\bibinfo {title} {Computational advantage from
  quantum-controlled ordering of gates},\ }\href
  {https://doi.org/10.1103/PhysRevLett.113.250402} {\bibfield  {journal}
  {\bibinfo  {journal} {Phys. Rev. Lett.}\ }\textbf {\bibinfo {volume} {113}},\
  \bibinfo {pages} {250402} (\bibinfo {year} {2014})}\BibitemShut {NoStop}%
\bibitem [{\citenamefont {Ebler}\ \emph {et~al.}(2018)\citenamefont {Ebler},
  \citenamefont {Salek},\ and\ \citenamefont
  {Chiribella}}]{PhysRevLett.120.120502}%
  \BibitemOpen
  \bibfield  {author} {\bibinfo {author} {\bibfnamefont {D.}~\bibnamefont
  {Ebler}}, \bibinfo {author} {\bibfnamefont {S.}~\bibnamefont {Salek}},\ and\
  \bibinfo {author} {\bibfnamefont {G.}~\bibnamefont {Chiribella}},\ }\bibfield
   {title} {\bibinfo {title} {Enhanced communication with the assistance of
  indefinite causal order},\ }\href
  {https://doi.org/10.1103/PhysRevLett.120.120502} {\bibfield  {journal}
  {\bibinfo  {journal} {Phys. Rev. Lett.}\ }\textbf {\bibinfo {volume} {120}},\
  \bibinfo {pages} {120502} (\bibinfo {year} {2018})}\BibitemShut {NoStop}%
\bibitem [{\citenamefont {{Salek}}\ \emph {et~al.}(2018)\citenamefont
  {{Salek}}, \citenamefont {{Ebler}},\ and\ \citenamefont
  {{Chiribella}}}]{2018arXiv180906655S}%
  \BibitemOpen
  \bibfield  {author} {\bibinfo {author} {\bibfnamefont {S.}~\bibnamefont
  {{Salek}}}, \bibinfo {author} {\bibfnamefont {D.}~\bibnamefont {{Ebler}}},\
  and\ \bibinfo {author} {\bibfnamefont {G.}~\bibnamefont {{Chiribella}}},\
  }\href@noop {} {\bibinfo {title} {{Quantum communication in a superposition
  of causal orders}}} (\bibinfo {year} {2018}),\ \Eprint
  {https://arxiv.org/abs/1809.06655} {arXiv:1809.06655 [quant-ph]} \BibitemShut
  {NoStop}%
\bibitem [{\citenamefont {Chiribella}\ \emph {et~al.}(2018)\citenamefont
  {Chiribella}, \citenamefont {Banik}, \citenamefont {Bhattacharya},
  \citenamefont {Guha}, \citenamefont {Alimuddin}, \citenamefont {Roy},
  \citenamefont {Saha}, \citenamefont {Agrawal},\ and\ \citenamefont
  {Kar}}]{chiribella2018indefinite}%
  \BibitemOpen
  \bibfield  {author} {\bibinfo {author} {\bibfnamefont {G.}~\bibnamefont
  {Chiribella}}, \bibinfo {author} {\bibfnamefont {M.}~\bibnamefont {Banik}},
  \bibinfo {author} {\bibfnamefont {S.~S.}\ \bibnamefont {Bhattacharya}},
  \bibinfo {author} {\bibfnamefont {T.}~\bibnamefont {Guha}}, \bibinfo {author}
  {\bibfnamefont {M.}~\bibnamefont {Alimuddin}}, \bibinfo {author}
  {\bibfnamefont {A.}~\bibnamefont {Roy}}, \bibinfo {author} {\bibfnamefont
  {S.}~\bibnamefont {Saha}}, \bibinfo {author} {\bibfnamefont {S.}~\bibnamefont
  {Agrawal}},\ and\ \bibinfo {author} {\bibfnamefont {G.}~\bibnamefont {Kar}},\
  }\href@noop {} {\bibinfo {title} {Indefinite causal order enables perfect
  quantum communication with zero capacity channel}} (\bibinfo {year} {2018}),\
  \Eprint {https://arxiv.org/abs/1810.10457} {arXiv:1810.10457 [quant-ph]}
  \BibitemShut {NoStop}%
\bibitem [{\citenamefont {{Procopio}}\ \emph {et~al.}(2019)\citenamefont
  {{Procopio}}, \citenamefont {{Delgado}}, \citenamefont {{Enriquez}},
  \citenamefont {{Belabas}},\ and\ \citenamefont
  {{Levenson}}}]{2019arXiv190201807P}%
  \BibitemOpen
  \bibfield  {author} {\bibinfo {author} {\bibfnamefont {L.~M.}\ \bibnamefont
  {{Procopio}}}, \bibinfo {author} {\bibfnamefont {F.}~\bibnamefont
  {{Delgado}}}, \bibinfo {author} {\bibfnamefont {M.}~\bibnamefont
  {{Enriquez}}}, \bibinfo {author} {\bibfnamefont {N.}~\bibnamefont
  {{Belabas}}},\ and\ \bibinfo {author} {\bibfnamefont {J.~A.}\ \bibnamefont
  {{Levenson}}},\ }\href@noop {} {\bibinfo {title} {{Communication through
  quantum coherent control of $N$ channels in a multi-partite causal-order
  scenario}}} (\bibinfo {year} {2019}),\ \Eprint
  {https://arxiv.org/abs/1902.01807} {arXiv:1902.01807 [quant-ph]} \BibitemShut
  {NoStop}%
\bibitem [{\citenamefont {{Mukhopadhyay}}\ \emph {et~al.}(2018)\citenamefont
  {{Mukhopadhyay}}, \citenamefont {{Gupta}},\ and\ \citenamefont
  {{Pati}}}]{2018arXiv181207508M}%
  \BibitemOpen
  \bibfield  {author} {\bibinfo {author} {\bibfnamefont {C.}~\bibnamefont
  {{Mukhopadhyay}}}, \bibinfo {author} {\bibfnamefont {M.~K.}\ \bibnamefont
  {{Gupta}}},\ and\ \bibinfo {author} {\bibfnamefont {A.~K.}\ \bibnamefont
  {{Pati}}},\ }\href@noop {} {\bibinfo {title} {{Superposition of causal order
  as a metrological resource for quantum thermometry}}} (\bibinfo {year}
  {2018}),\ \Eprint {https://arxiv.org/abs/1812.07508} {arXiv:1812.07508
  [quant-ph]} \BibitemShut {NoStop}%
\bibitem [{\citenamefont {Zhao}\ \emph {et~al.}(2019)\citenamefont {Zhao},
  \citenamefont {Yang},\ and\ \citenamefont {Chiribella}}]{zhao2019quantum}%
  \BibitemOpen
  \bibfield  {author} {\bibinfo {author} {\bibfnamefont {X.}~\bibnamefont
  {Zhao}}, \bibinfo {author} {\bibfnamefont {Y.}~\bibnamefont {Yang}},\ and\
  \bibinfo {author} {\bibfnamefont {G.}~\bibnamefont {Chiribella}},\
  }\href@noop {} {\bibinfo {title} {Quantum metrology with indefinite causal
  order}} (\bibinfo {year} {2019}),\ \Eprint {https://arxiv.org/abs/1912.02449}
  {arXiv:1912.02449 [quant-ph]} \BibitemShut {NoStop}%
\bibitem [{\citenamefont {Frey}(2019)}]{DBLP:journals/qip/Frey19}%
  \BibitemOpen
  \bibfield  {author} {\bibinfo {author} {\bibfnamefont {M.}~\bibnamefont
  {Frey}},\ }\bibfield  {title} {\bibinfo {title} {Indefinite causal order aids
  quantum depolarizing channel identification},\ }\href
  {https://doi.org/10.1007/s11128-019-2186-9} {\bibfield  {journal} {\bibinfo
  {journal} {Quantum Inf. Process.}\ }\textbf {\bibinfo {volume} {18}},\
  \bibinfo {pages} {96} (\bibinfo {year} {2019})}\BibitemShut {NoStop}%
\bibitem [{\citenamefont {Chapeau-Blondeau}(2021)}]{PhysRevA.103.032615}%
  \BibitemOpen
  \bibfield  {author} {\bibinfo {author} {\bibfnamefont {F.~m.~c.}\
  \bibnamefont {Chapeau-Blondeau}},\ }\bibfield  {title} {\bibinfo {title}
  {Noisy quantum metrology with the assistance of indefinite causal order},\
  }\href {https://doi.org/10.1103/PhysRevA.103.032615} {\bibfield  {journal}
  {\bibinfo  {journal} {Phys. Rev. A}\ }\textbf {\bibinfo {volume} {103}},\
  \bibinfo {pages} {032615} (\bibinfo {year} {2021})}\BibitemShut {NoStop}%
\bibitem [{\citenamefont {Procopio}\ \emph {et~al.}(2015)\citenamefont
  {Procopio}, \citenamefont {Moqanaki}, \citenamefont {Ara{\'{u}}jo},
  \citenamefont {Costa}, \citenamefont {Calafell}, \citenamefont {Dowd},
  \citenamefont {Hamel}, \citenamefont {Rozema}, \citenamefont {Brukner},\ and\
  \citenamefont {Walther}}]{Procopio2015}%
  \BibitemOpen
  \bibfield  {author} {\bibinfo {author} {\bibfnamefont {L.~M.}\ \bibnamefont
  {Procopio}}, \bibinfo {author} {\bibfnamefont {A.}~\bibnamefont {Moqanaki}},
  \bibinfo {author} {\bibfnamefont {M.}~\bibnamefont {Ara{\'{u}}jo}}, \bibinfo
  {author} {\bibfnamefont {F.}~\bibnamefont {Costa}}, \bibinfo {author}
  {\bibfnamefont {I.~A.}\ \bibnamefont {Calafell}}, \bibinfo {author}
  {\bibfnamefont {E.~G.}\ \bibnamefont {Dowd}}, \bibinfo {author}
  {\bibfnamefont {D.~R.}\ \bibnamefont {Hamel}}, \bibinfo {author}
  {\bibfnamefont {L.~A.}\ \bibnamefont {Rozema}}, \bibinfo {author}
  {\bibfnamefont {{\v{C}}.}~\bibnamefont {Brukner}},\ and\ \bibinfo {author}
  {\bibfnamefont {P.}~\bibnamefont {Walther}},\ }\bibfield  {title} {\bibinfo
  {title} {Experimental superposition of orders of quantum gates},\ }\bibfield
  {journal} {\bibinfo  {journal} {Nature Communications}\ }\textbf {\bibinfo
  {volume} {6}},\ \href {https://doi.org/10.1038/ncomms8913}
  {10.1038/ncomms8913} (\bibinfo {year} {2015})\BibitemShut {NoStop}%
\bibitem [{\citenamefont {Rubino}\ \emph {et~al.}(2017)\citenamefont {Rubino},
  \citenamefont {Rozema}, \citenamefont {Feix}, \citenamefont {Ara{\'{u}}jo},
  \citenamefont {Zeuner}, \citenamefont {Procopio}, \citenamefont {Brukner},\
  and\ \citenamefont {Walther}}]{rubino2017}%
  \BibitemOpen
  \bibfield  {author} {\bibinfo {author} {\bibfnamefont {G.}~\bibnamefont
  {Rubino}}, \bibinfo {author} {\bibfnamefont {L.~A.}\ \bibnamefont {Rozema}},
  \bibinfo {author} {\bibfnamefont {A.}~\bibnamefont {Feix}}, \bibinfo {author}
  {\bibfnamefont {M.}~\bibnamefont {Ara{\'{u}}jo}}, \bibinfo {author}
  {\bibfnamefont {J.~M.}\ \bibnamefont {Zeuner}}, \bibinfo {author}
  {\bibfnamefont {L.~M.}\ \bibnamefont {Procopio}}, \bibinfo {author}
  {\bibfnamefont {{\v{C}}.}~\bibnamefont {Brukner}},\ and\ \bibinfo {author}
  {\bibfnamefont {P.}~\bibnamefont {Walther}},\ }\bibfield  {title} {\bibinfo
  {title} {Experimental verification of an indefinite causal order},\ }\href
  {https://doi.org/10.1126/sciadv.1602589} {\bibfield  {journal} {\bibinfo
  {journal} {Science Advances}\ }\textbf {\bibinfo {volume} {3}},\ \bibinfo
  {pages} {e1602589} (\bibinfo {year} {2017})}\BibitemShut {NoStop}%
\bibitem [{\citenamefont {Goswami}\ \emph
  {et~al.}(2018{\natexlab{a}})\citenamefont {Goswami}, \citenamefont
  {Giarmatzi}, \citenamefont {Kewming}, \citenamefont {Costa}, \citenamefont
  {Branciard}, \citenamefont {Romero},\ and\ \citenamefont
  {White}}]{PhysRevLett.121.090503}%
  \BibitemOpen
  \bibfield  {author} {\bibinfo {author} {\bibfnamefont {K.}~\bibnamefont
  {Goswami}}, \bibinfo {author} {\bibfnamefont {C.}~\bibnamefont {Giarmatzi}},
  \bibinfo {author} {\bibfnamefont {M.}~\bibnamefont {Kewming}}, \bibinfo
  {author} {\bibfnamefont {F.}~\bibnamefont {Costa}}, \bibinfo {author}
  {\bibfnamefont {C.}~\bibnamefont {Branciard}}, \bibinfo {author}
  {\bibfnamefont {J.}~\bibnamefont {Romero}},\ and\ \bibinfo {author}
  {\bibfnamefont {A.~G.}\ \bibnamefont {White}},\ }\bibfield  {title} {\bibinfo
  {title} {Indefinite causal order in a quantum switch},\ }\href
  {https://doi.org/10.1103/PhysRevLett.121.090503} {\bibfield  {journal}
  {\bibinfo  {journal} {Phys. Rev. Lett.}\ }\textbf {\bibinfo {volume} {121}},\
  \bibinfo {pages} {090503} (\bibinfo {year} {2018}{\natexlab{a}})}\BibitemShut
  {NoStop}%
\bibitem [{\citenamefont {Goswami}\ \emph
  {et~al.}(2018{\natexlab{b}})\citenamefont {Goswami}, \citenamefont {Cao},
  \citenamefont {Paz-Silva}, \citenamefont {Romero},\ and\ \citenamefont
  {White}}]{goswami2018communicating}%
  \BibitemOpen
  \bibfield  {author} {\bibinfo {author} {\bibfnamefont {K.}~\bibnamefont
  {Goswami}}, \bibinfo {author} {\bibfnamefont {Y.}~\bibnamefont {Cao}},
  \bibinfo {author} {\bibfnamefont {G.~A.}\ \bibnamefont {Paz-Silva}}, \bibinfo
  {author} {\bibfnamefont {J.}~\bibnamefont {Romero}},\ and\ \bibinfo {author}
  {\bibfnamefont {A.~G.}\ \bibnamefont {White}},\ }\href@noop {} {\bibinfo
  {title} {Communicating via ignorance}} (\bibinfo {year}
  {2018}{\natexlab{b}}),\ \Eprint {https://arxiv.org/abs/1807.07383}
  {arXiv:1807.07383 [quant-ph]} \BibitemShut {NoStop}%
\bibitem [{\citenamefont {Wei}\ \emph {et~al.}(2019)\citenamefont {Wei},
  \citenamefont {Tischler}, \citenamefont {Zhao}, \citenamefont {Li},
  \citenamefont {Arrazola}, \citenamefont {Liu}, \citenamefont {Zhang},
  \citenamefont {Li}, \citenamefont {You}, \citenamefont {Wang}, \citenamefont
  {Chen}, \citenamefont {Sanders}, \citenamefont {Zhang}, \citenamefont
  {Pryde}, \citenamefont {Xu},\ and\ \citenamefont
  {Pan}}]{PhysRevLett.122.120504}%
  \BibitemOpen
  \bibfield  {author} {\bibinfo {author} {\bibfnamefont {K.}~\bibnamefont
  {Wei}}, \bibinfo {author} {\bibfnamefont {N.}~\bibnamefont {Tischler}},
  \bibinfo {author} {\bibfnamefont {S.-R.}\ \bibnamefont {Zhao}}, \bibinfo
  {author} {\bibfnamefont {Y.-H.}\ \bibnamefont {Li}}, \bibinfo {author}
  {\bibfnamefont {J.~M.}\ \bibnamefont {Arrazola}}, \bibinfo {author}
  {\bibfnamefont {Y.}~\bibnamefont {Liu}}, \bibinfo {author} {\bibfnamefont
  {W.}~\bibnamefont {Zhang}}, \bibinfo {author} {\bibfnamefont
  {H.}~\bibnamefont {Li}}, \bibinfo {author} {\bibfnamefont {L.}~\bibnamefont
  {You}}, \bibinfo {author} {\bibfnamefont {Z.}~\bibnamefont {Wang}}, \bibinfo
  {author} {\bibfnamefont {Y.-A.}\ \bibnamefont {Chen}}, \bibinfo {author}
  {\bibfnamefont {B.~C.}\ \bibnamefont {Sanders}}, \bibinfo {author}
  {\bibfnamefont {Q.}~\bibnamefont {Zhang}}, \bibinfo {author} {\bibfnamefont
  {G.~J.}\ \bibnamefont {Pryde}}, \bibinfo {author} {\bibfnamefont
  {F.}~\bibnamefont {Xu}},\ and\ \bibinfo {author} {\bibfnamefont {J.-W.}\
  \bibnamefont {Pan}},\ }\bibfield  {title} {\bibinfo {title} {Experimental
  quantum switching for exponentially superior quantum communication
  complexity},\ }\href {https://doi.org/10.1103/PhysRevLett.122.120504}
  {\bibfield  {journal} {\bibinfo  {journal} {Phys. Rev. Lett.}\ }\textbf
  {\bibinfo {volume} {122}},\ \bibinfo {pages} {120504} (\bibinfo {year}
  {2019})}\BibitemShut {NoStop}%
\bibitem [{\citenamefont {Guo}\ \emph {et~al.}(2020)\citenamefont {Guo},
  \citenamefont {Hu}, \citenamefont {Hou}, \citenamefont {Cao}, \citenamefont
  {Cui}, \citenamefont {Liu}, \citenamefont {Huang}, \citenamefont {Li},
  \citenamefont {Guo},\ and\ \citenamefont
  {Chiribella}}]{PhysRevLett.124.030502}%
  \BibitemOpen
  \bibfield  {author} {\bibinfo {author} {\bibfnamefont {Y.}~\bibnamefont
  {Guo}}, \bibinfo {author} {\bibfnamefont {X.-M.}\ \bibnamefont {Hu}},
  \bibinfo {author} {\bibfnamefont {Z.-B.}\ \bibnamefont {Hou}}, \bibinfo
  {author} {\bibfnamefont {H.}~\bibnamefont {Cao}}, \bibinfo {author}
  {\bibfnamefont {J.-M.}\ \bibnamefont {Cui}}, \bibinfo {author} {\bibfnamefont
  {B.-H.}\ \bibnamefont {Liu}}, \bibinfo {author} {\bibfnamefont {Y.-F.}\
  \bibnamefont {Huang}}, \bibinfo {author} {\bibfnamefont {C.-F.}\ \bibnamefont
  {Li}}, \bibinfo {author} {\bibfnamefont {G.-C.}\ \bibnamefont {Guo}},\ and\
  \bibinfo {author} {\bibfnamefont {G.}~\bibnamefont {Chiribella}},\ }\bibfield
   {title} {\bibinfo {title} {Experimental transmission of quantum information
  using a superposition of causal orders},\ }\href
  {https://doi.org/10.1103/PhysRevLett.124.030502} {\bibfield  {journal}
  {\bibinfo  {journal} {Phys. Rev. Lett.}\ }\textbf {\bibinfo {volume} {124}},\
  \bibinfo {pages} {030502} (\bibinfo {year} {2020})}\BibitemShut {NoStop}%
\bibitem [{\citenamefont {Felce}\ and\ \citenamefont
  {Vedral}(2020)}]{ICOFridgeTheory}%
  \BibitemOpen
  \bibfield  {author} {\bibinfo {author} {\bibfnamefont {D.}~\bibnamefont
  {Felce}}\ and\ \bibinfo {author} {\bibfnamefont {V.}~\bibnamefont {Vedral}},\
  }\bibfield  {title} {\bibinfo {title} {Quantum refrigeration with indefinite
  causal order},\ }\href {https://doi.org/10.1103/PhysRevLett.125.070603}
  {\bibfield  {journal} {\bibinfo  {journal} {Phys. Rev. Lett.}\ }\textbf
  {\bibinfo {volume} {125}},\ \bibinfo {pages} {070603} (\bibinfo {year}
  {2020})}\BibitemShut {NoStop}%
\bibitem [{\citenamefont {Nie}\ \emph {et~al.}(2020)\citenamefont {Nie},
  \citenamefont {Zhu}, \citenamefont {Xi}, \citenamefont {Long}, \citenamefont
  {Lin}, \citenamefont {Tian}, \citenamefont {Qiu}, \citenamefont {Yang},
  \citenamefont {Dong}, \citenamefont {Li}, \citenamefont {Xin},\ and\
  \citenamefont {Lu}}]{nie2020experimental}%
  \BibitemOpen
  \bibfield  {author} {\bibinfo {author} {\bibfnamefont {X.}~\bibnamefont
  {Nie}}, \bibinfo {author} {\bibfnamefont {X.}~\bibnamefont {Zhu}}, \bibinfo
  {author} {\bibfnamefont {C.}~\bibnamefont {Xi}}, \bibinfo {author}
  {\bibfnamefont {X.}~\bibnamefont {Long}}, \bibinfo {author} {\bibfnamefont
  {Z.}~\bibnamefont {Lin}}, \bibinfo {author} {\bibfnamefont {Y.}~\bibnamefont
  {Tian}}, \bibinfo {author} {\bibfnamefont {C.}~\bibnamefont {Qiu}}, \bibinfo
  {author} {\bibfnamefont {X.}~\bibnamefont {Yang}}, \bibinfo {author}
  {\bibfnamefont {Y.}~\bibnamefont {Dong}}, \bibinfo {author} {\bibfnamefont
  {J.}~\bibnamefont {Li}}, \bibinfo {author} {\bibfnamefont {T.}~\bibnamefont
  {Xin}},\ and\ \bibinfo {author} {\bibfnamefont {D.}~\bibnamefont {Lu}},\
  }\href@noop {} {\bibinfo {title} {Experimental realization of a quantum
  refrigerator driven by indefinite causal orders}} (\bibinfo {year} {2020}),\
  \Eprint {https://arxiv.org/abs/2011.12580} {arXiv:2011.12580 [quant-ph]}
  \BibitemShut {NoStop}%
\bibitem [{\citenamefont {Cao}\ \emph {et~al.}(2021)\citenamefont {Cao},
  \citenamefont {ning Wang}, \citenamefont {Jia}, \citenamefont {Zhang},
  \citenamefont {Guo}, \citenamefont {Liu}, \citenamefont {Huang},
  \citenamefont {Li},\ and\ \citenamefont {Guo}}]{cao2021experimental}%
  \BibitemOpen
  \bibfield  {author} {\bibinfo {author} {\bibfnamefont {H.}~\bibnamefont
  {Cao}}, \bibinfo {author} {\bibfnamefont {N.}~\bibnamefont {ning Wang}},
  \bibinfo {author} {\bibfnamefont {Z.-A.}\ \bibnamefont {Jia}}, \bibinfo
  {author} {\bibfnamefont {C.}~\bibnamefont {Zhang}}, \bibinfo {author}
  {\bibfnamefont {Y.}~\bibnamefont {Guo}}, \bibinfo {author} {\bibfnamefont
  {B.-H.}\ \bibnamefont {Liu}}, \bibinfo {author} {\bibfnamefont {Y.-F.}\
  \bibnamefont {Huang}}, \bibinfo {author} {\bibfnamefont {C.-F.}\ \bibnamefont
  {Li}},\ and\ \bibinfo {author} {\bibfnamefont {G.-C.}\ \bibnamefont {Guo}},\
  }\href@noop {} {\bibinfo {title} {Experimental demonstration of indefinite
  causal order induced quantum heat extraction}} (\bibinfo {year} {2021}),\
  \Eprint {https://arxiv.org/abs/2101.07979} {arXiv:2101.07979 [quant-ph]}
  \BibitemShut {NoStop}%
\bibitem [{\citenamefont {Nielsen}\ and\ \citenamefont
  {Chuang}(2000)}]{NielsenChuang}%
  \BibitemOpen
  \bibfield  {author} {\bibinfo {author} {\bibfnamefont {M.~A.}\ \bibnamefont
  {Nielsen}}\ and\ \bibinfo {author} {\bibfnamefont {I.~L.}\ \bibnamefont
  {Chuang}},\ }\href@noop {} {\emph {\bibinfo {title} {Quantum Computation and
  Quantum Information}}}\ (\bibinfo  {publisher} {Cambridge University Press},\
  \bibinfo {address} {New York, USA},\ \bibinfo {year} {2000})\ p.\ \bibinfo
  {pages} {360}\BibitemShut {NoStop}%
\end{thebibliography}%

\end{document}